\documentclass[reprint,showpacs,preprintnumbers,amsmath,amssymb, aps]{revtex4-1}
\usepackage{graphicx,wasysym}
\usepackage{dcolumn}
\usepackage{bm}
\usepackage{longtable}
\usepackage{epstopdf}
\setlongtables

\usepackage{tabulary}

\newcommand{\Tm}{T_{\mathrm{m}}}
\newcommand{\Tg}{T_{\mathrm{g}}}
\newcommand{\Ts}{T_{\mathrm{s}}}
\newcommand{\To}{T_{\mathrm{o}}}
\newcommand{\Tx}{T_{\mathrm{max}}}

\newcommand{\po}{p_{\mathrm{o}}}

\newcommand{\taul}{\tau_{\mathrm{liq}}}

\newcommand{\An}{\mathrm{\AA}}

\begin{document}

\preprint{1}

\title{Corresponding states for mesostructure and dynamics of supercooled water}


\author{David T. Limmer and David Chandler}

 \email{chandler@berkeley.edu}
\affiliation{%
Department of Chemistry, University of California, Berkeley, CA, USA 94609
}%
\date{\today}
\begin{abstract}

Water famously expands upon freezing, foreshadowed by a negative coefficient of expansion of the liquid at temperatures close to its freezing temperature. These behaviors, and many others, reflect the energetic preference for local tetrahedral arrangements of water molecules and entropic effects that oppose it. Here, we provide theoretical analysis of mesoscopic implications of this competition, both equilibrium and non-equilibrium, including mediation by interfaces. With general scaling arguments bolstered by simulation results, and with reduced units that elucidate corresponding states, we derive a phase diagram for bulk and confined water and water-like materials.  For water itself, the corresponding states cover the temperature range of 150\,K to 300\,K and the pressure range of 1\,bar to 2\,kbar.  In this regime, there are two reversible condensed phases --  ice and liquid. Out of equilibrium, there is irreversible polyamorphism, i.e., more than one glass phase, reflecting dynamical arrest of coarsening ice. Temperature-time plots are derived to characterize time scales of the different phases and explain contrasting dynamical behaviors of different water-like systems. 
\end{abstract}

\maketitle
\section{Introduction}

Supercooled liquids exist in a metastable equilibrium made possible by a separation of timescales between local liquid equilibration and global crystallization.\cite{debenedetti1996metastable} Supercooled water is no different in this regard. However, the magnitude of the separation of timescales in supercooled water is of particular relevance due to speculation regarding the behavior of the thermodynamic properties of liquid water at very low temperatures.\cite{angell1983supercooled} In this work, we describe a theory for corresponding states that relates the low temperature, low pressure phase diagram with the time-temperature-transformation diagram for supercooled water and water-like systems. Our derivations use scaling theories with assumptions tested against molecular simulation.  The relationships elucidate the connections between behaviors found for different molecular simulation models of water and for different water-like substances.

\begin{figure}[b]
\begin{center}
\includegraphics[width=8.5cm]{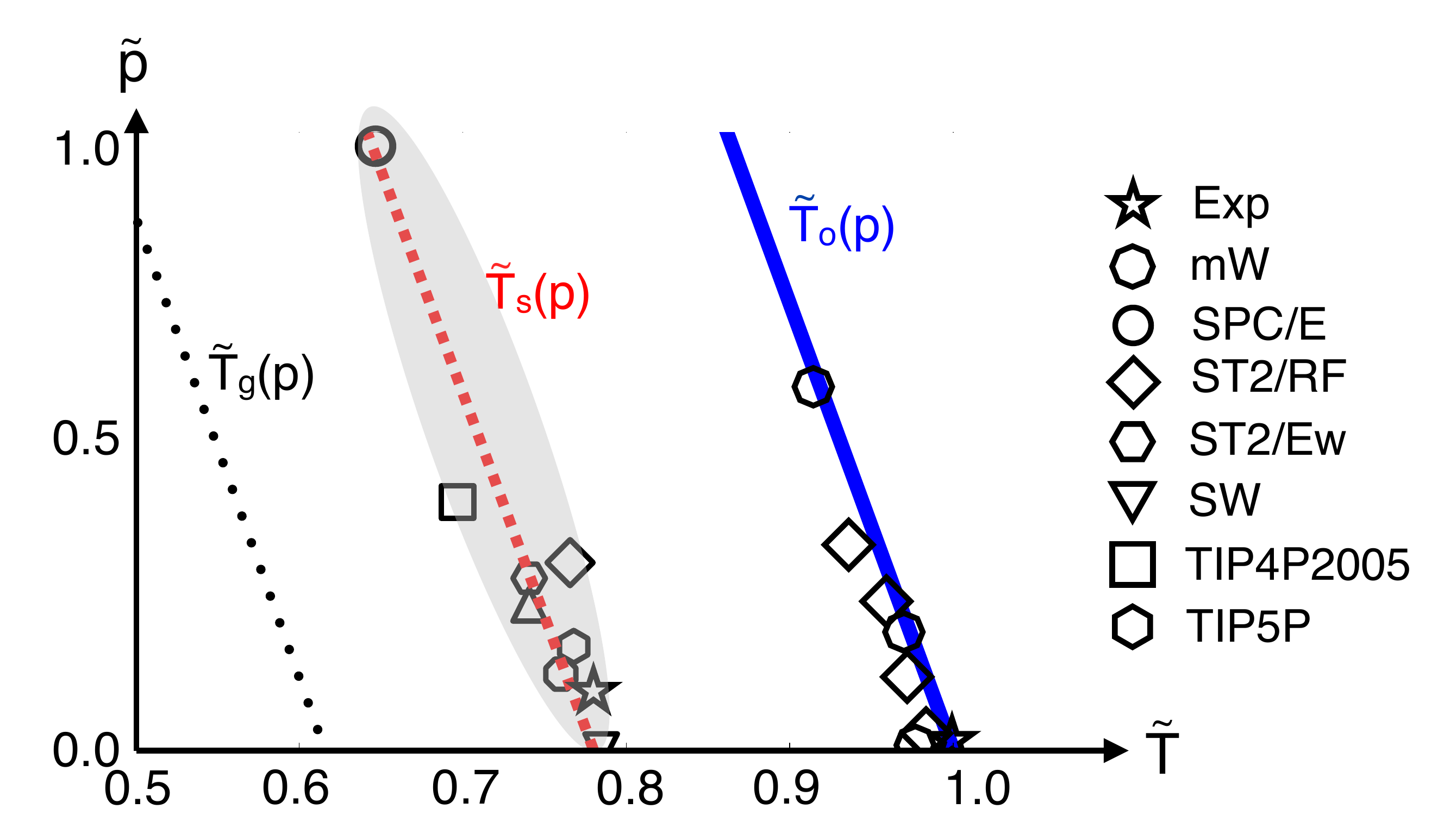}
\caption{\label{fig1}The $\tilde{p}$-$\tilde{T}$ phase diagram for supercooled water.  These symbols refer to the pressure, $p$, and the temperature $T$, in units of the reference pressure, $\po$, and the reference temperature, $\Tx$, for each specific material.\cite{Stillinger:1985p5262, Stillinger:1974p1545, vega2005relation, Molinero:2009p4008, abascal2005general}  See text.  The lines refer to theoretically expected trends as functions of pressure for the dynamical onset temperature, $\To$, for the liquid limit of stability temperature, $\Ts$, and for a glass transition temperature, $\Tg$.  Symbols indicate locations of these these temperatures in the reduced units as obtained from experiment and from different molecular simulation models, where the key employs standard acronyms for each system.\cite{abascal2010widom, poole2013free, liu2009low, brovchenko2003multiple, vasisht2011liquid, xu2011there}  For the glass transition, the line marks the stage where the liquid reorganization time exceeds $10^{14}\,\tau_\mathrm{o}$, where the reference time, $\tau_\mathrm{o}$, is that reorganization time at the onset to correlated dynamics ($\tau_\mathrm{o} \approx 1\,\mathrm{ps}$ for liquid water).}
\end{center} 
\end{figure}

Figure~\ref{fig1} shows the portion of the phase diagram for supercooled water relevant to this paper.  Temperature, $T$, ranges from ambient conditions to deep into the supercooled regime, and pressure, $p$, ranges from atmospheric conditions through the range of stability for ordinary hexagonal ice. Experimentally, this region corresponds to  $150~\mathrm{K}< T < 300~\mathrm{K}$ and $0~\mathrm{kbar}< p  < 2~\mathrm{kbar}$.\cite{eisenberg2005structure} The locations of specific features relative to experiment  vary from one molecular model to another.\cite{sanz2004phase} This variability reflects a delicate competition between entropy and energy that is intrinsic to any reasonable model of water or water-like system. 

One manifestation of this competition is the existence of the temperature of maximum density. We use $\Tx$ to denote the value of this temperature at ambient (i.e., low pressure) conditions.  For experimental water, $\Tx \approx 277$\,K.  The energy-entropy balance manifested in the density maximum is shifted to lower temperatures as elevated pressures favor denser packing.  A measure of this shift is provided by the slope of the melting line or in terms of a reference pressure $\po = - \Delta H/10\, \Delta V $, where $\Delta H$ and $\Delta V$ are, respectively, the enthalpy and volume changes upon melting. For experimental water, $\po \approx 3.7$\,kbar.  We use $\Tx$ and $\po$ to compare the properties of different water models as well as to enable comparison with experiment.\cite{Stillinger:1985p5262, Stillinger:1974p1545, vega2005relation, Molinero:2009p4008, abascal2005general,broughton1987phase,eisenberg2005structure,abascal2005general}  Thus, the phase diagram in Fig.~\ref{fig1} employs the reduced variables
\begin{equation}
\label{reduced}
\tilde{p} = p/\po \quad \mathrm{and} \quad \tilde{T} = T/\Tx\,.
\end{equation}

In this way, Fig.~\ref{fig1} relates results from various models and experiments for the onset temperature, $\To$, and the homogeneous nucleation temperature, $\Ts$.  The former, $\To$, marks the crossover to correlated (i.e., hierarchical) dynamics.\cite{chandler2010dynamics} The latter, $\Ts$, marks the crossover to liquid instability.\cite{Limmer:2011p134503,limmer2012phase,limmer2013putative}  These temperatures are material properties.  Figure~\ref{fig1} also shows a reduced glass transition temperature, $\tilde{\Tg} = \Tg/\Tx$.  This temperature is defined as that where the reversible structural relaxation time of liquid water equals 100\,s.  

Glass phases of water, where aging occurs on time scales of 100\,s or longer, are not generally accessible by straightforward supercooling because bulk liquid water spontaneously freezes into crystal ice at temperatures below $\Ts$.  Freezing in this regime occurs in mili-second or shorter time scales.\cite{koop:2000p611}  An amorphous solid can be reached with a cooling trajectory that is initially fast enough to arrest crystallization, and finally cold enough to produce very slow aging.  Alternatively, one may cool while perturbing water with surfaces that inhibit crystallization.  An actual glass transition temperature of water is therefore not a material property because its value depends upon the protocol by which the material is driven out of equilibrium. Surface mediated approaches to amorphous solids can yield $\Tg$'s that are higher than those produced by rapid temperature quenches.  The $\Tg$ graphed in Fig.~\ref{fig1} is necessarily an upper bound to those glass transition temperatures.  

Dynamics in the vicinity of $T\approx\Ts$ exhibits a two-step coarsening of the crystal phase.\cite{limmer2013putative,moore2011structural,moore2010ice} First, disperse nano-scale domains of local crystal order form throughout the melt; second, on a much longer time scale, the nano-scale domains meld into much larger ordered domains.  These steps are arrested when forming glass.\cite{limmer2013amorphous}  Configurations appearing at the initial stages of this coarsening are often observed in computer simulations of water.  These configurations are transient states that are almost as often confused with the presence of two distinct supercooled liquid phases,\footnote{The list of representative papers is long.  We have provided a summary elsewhere.\cite{limmer2013putative}} and claims that some water-like models do not exhibit this behavior\cite{giovambattista2012interplay} are based upon studies that have not examined this part of the phase diagram.  Two distinct reversible liquids in coexistence would imply the existence of a low temperature critical point of the sort suggested by Stanley and his co-workers.\cite{Poole:1992p324} Not surprisingly, all reports of a low-temperature critical point in water or water-like systems locate a point on or near $\Ts(p)$.  In fact, all the simulation points clustered around that line in Fig.~\ref{fig1} have been incorrectly identified as low-temperature critical points.\cite{abascal2010widom, poole2013free, liu2009low, brovchenko2003multiple, vasisht2011liquid, xu2011there}  Similarly, in experimental work, Mishima locates a putative critical point\cite{mishima2010volume}  close to the experimental limit of liquid stability.\cite{speedy1976isothermal}  We have analyzed and disproved this notion of a liquid-liquid transition for several different computer simulation models of water and water-like systems.\cite{limmer2013putative}  

We mention the disproved notion only to emphasize that the equilibrium phase diagram by itself gives an incomplete picture of the behavior of supercooled water.  Supercooled water, being a metastable state, behaves reversibly for only finite observation times. The specific length of that time depends on the separation of timescales between local equilibration of liquid configurations and global crystallization. When the gap between these timescales becomes small, as it does for either $T < \Ts$ or $T \approx \Ts$, time-independent thermodynamic properties are no longer well defined.\\

\section{Universal temperature dependence of liquid relaxation times for supercooled water}

\begin{figure*}
\begin{center}
\includegraphics[width=13cm]{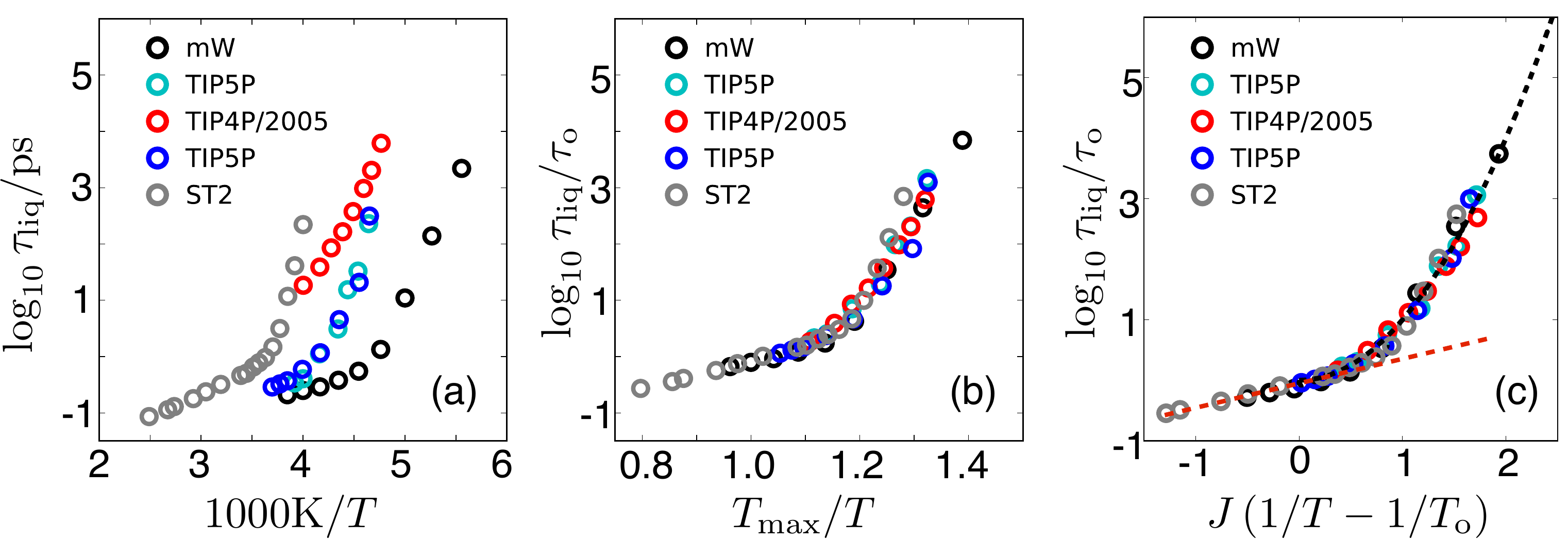}
\caption{Low temperature variation of structural relaxation time, $\taul$, for different models of water. (a) Logarithm of that time as a function of absolute temperature $T$. (b) The same data, now in units of the reference time $\tau_\mathrm{o}$, and as a function of reduced temperature $\tilde{T} = T/\Tx$. The reference time, $\tau_\mathrm{o}$, is the structural relaxation time at the onset temperature, $\To$.  (c) The same data, now collapsed to the parabolic form, Eq.~\ref{Eq:parabola}, where the dashed black line is the prediction of that equation, and the dashed red line is the Arrhenius form that holds for temperatures above $\To$.  }
\label{Fi:tau_model}
\end{center} 
\end{figure*}

In order to construct a scaling theory for the liquid relaxation time, $\taul$, we follow our previous work\cite{limmer2012phase} in adopting a perspective of dynamic facilitation theory.\cite{chandler2010dynamics}  An important aspect of this perspective is that it supplies a universal form for the relaxation time as a function of temperature. This form, known as the ``parabolic law'', is
\begin{equation}\label{Eq:parabola}
\log_{10} (\taul/\tau_\mathrm{o}) = J^2 \, \left (1/T - 1/\To \right)^2 \,\quad \mathrm{for} \, \quad T<\To\, ,
\end{equation}
where $J$ is an energy scale of hierarchical dynamics, $\To$ is the temperature below which that dynamics sets in, and $\tau_\mathrm{o}$ is the liquid relaxation time at the onset temperature $\To$.\footnote{In general, $\tau_\mathrm{o}$ may itself multiply an Arhrenius temperature dependent factor, but we neglect this quantitative detail here because it is a small effect compared to the super-Arrhenius behavior at low temperature.} This form has been used to collapse large and seemingly disparate collections of experimental and simulation data.\cite{elmatad2010corresponding} Figure~\ref{Fi:tau_model} illustrates the nature of this collapse for the structural relaxation times of several models of water.\cite{wikfeldt2011enhanced,yamada2002interplay,poole2011dynamical} 

For the data shown in Fig.~\ref{Fi:tau_model}, the relaxation times have been calculated from the self-correlation function,
\begin{equation}\label{eq:fskt}
F_\mathrm{s}(k,t) = \left < \mathrm{e}^{i \mathbf{k}\cdot \left[ \mathbf{r}_1(t) - \mathbf{r}_1(0) \right]} \right > \, ,
\end{equation}
for wave vectors of magnitude $k\approx 2\pi/ 2.8 \An$. Here, $\mathbf{r}_1(t)$ denotes the position of a tagged molecule at time $t$, and the angle brackets stand for the equilibrium average over initial conditions.  The time at which this $F_\mathrm{s}(k,t)$ decays to 1/e of its initial value is defined as $\taul$. In all of the models studied, the temperature dependence of this time crosses over from a weak Arrhenius temperature dependence to a super-Arrhenius temperature dependence. The location where the crossover occurs, the onset temperature $T_\mathrm{o}$, varies from model to model as does the reference timescale $\tau_\mathrm{o}$. This variability in part reflects quantitative differences between phase diagrams for each of the models.  Indeed, Fig.~\ref{Fi:tau_model}b shows that the temperature dependence of the relaxation times can be collapsed by referencing the data to the temperature of maximum density.  It is a remarkable result given the wide variation of $\Tx$, ranging from 250\,K to 320\,K for the different models.\cite{vega2005relation} 

In addition, Fig.~\ref{Fi:tau_model}c shows that the relaxation time data also collapses when referenced to the onset temperature, $\To$, and that the collapsed data obeys the parabolic law for all $T<\To$.  This finding establishes that  
\begin{equation}
\label{ToTmax}
\To \approx \Tx \, .
\end{equation}
Further, collapsing data to the parabolic law reveals that $J/\To$ varies between models of water by no more that  5\% and on average by only 1\%.  This universal value is $J/\To \approx 7.4$. For the models considered here,  $\tau_\mathrm{o}$ varies between 0.3\,ps and 8.0\,ps, which largely reflects the density differences between models at low pressure. 

The collapse of the relaxation times for the difference models as a function of $\Tx$ implies a universality in the behavior of the glass transition for these models and, by proxy, for experiment. As we have done previously,\cite{limmer2012phase} we can define a locus of laboratory glass transitions as the locations in the phase diagram where the liquid relaxation time is equal to $10^{14}\, \tau_\mathrm{o}$, which for many of the models implies $\taul(\Tg) \approx 100$\,s, so that with Eq.~\ref{Eq:parabola} we have,
\begin{equation}
\label{eq:Tg}
\Tg/\To \approx \left ( \sqrt{14\,}\, \To/J +1 \right )^{-1} \, .
\end{equation}
Taking $J/\To \approx 7.4$ and $\To \approx \Tx$, we therefore conclude that for water and water-like models, $\Tg \approx 0.62\, \Tx$. This value yields the glass-transition line graphed in Fig.~\ref{fig1}, where the slope of the line is the same as that for $\To(p)$. Experimentally, the density maximum for water occurs at $\Tx \approx 277$, therefore our predicted glass transition is $\Tg \approx 172$~K.  This temperature agrees with our previous work inferring the glass transition temperature from relaxation data of confined water.\cite{limmer2012phase}  It also provides an upper bound to values for $\Tg$ obtained with other experimental protocols.\cite{capaccioli2011resolving,angell2002liquid} 

\section{Molecular theory for $J$, $\tau_\mathrm{o}$ and $\To$}

The energy, time and temperature parameters in Eq.~\ref{Eq:parabola} can be computed from microscopic theory following the procedures of Keys et. al.\cite{keys2011excitations}  The procedures are based upon mapping the dynamics of atomic degrees of freedom to dynamics of a kinetically constrained East model.\cite{jackle1991hierarchically}  The parabolic law, Eq.~\ref{Eq:parabola}, is a consequence of that mapping.  

To illustrate the procedure for water, we have carried out molecular dynamics simulations of equilibrated water models to determine the net number of enduring displacements of length $a$ appearing in $N$-molecule trajectories that run for observation times $t_\mathrm{obs}$.  This number of displacements is 
\begin{equation}\label{eq:exite}
C_a = \sum_{i=1}^N \sum_{j=0}^{t_\mathrm{obs}/\Delta t} \Theta\left(|\bar{\mathbf{r}}_i(j\Delta t + \Delta t )-\bar{\mathbf{r}}_i(j\Delta t ) | - a \right )
\end{equation}
where $\Theta(x)$ is 1 for $x>0$ and zero otherwise, $\Delta t$ is the mean instanton time for enduring displacements of length $a$, and $\bar{\mathbf{r}}_i(t)$ is the position of particle $i$ averaged over the time interval $t-\delta t/2$ to $t+\delta t/2$.  The averaging over $\delta t$ coarse-grains out irrelevant vibrational motions.  The instanton time, $\Delta t$, is taken to be large enough that non-enduring transitions are also removed from consideration.  The two times, $\Delta t > \delta t$, are determined as prescribed by Keys et. al.\cite{keys2011excitations}  

\begin{figure}
\begin{center}
\includegraphics[width=7.5cm]{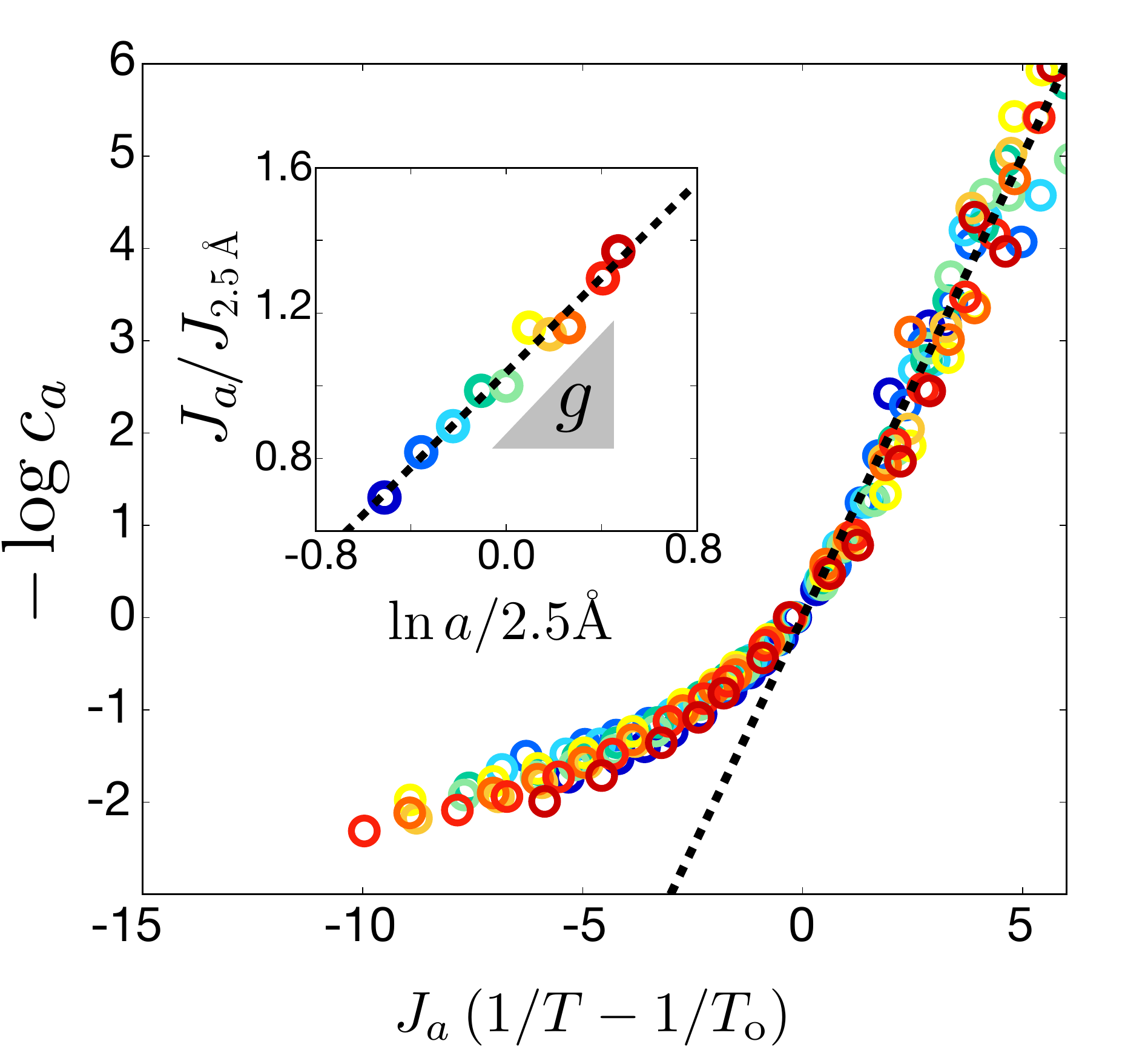}
\caption{Excitation concentration for the mW model at ambient pressure for enduring displacement lengthscales, $a$, between 1.5 and 3.5 $\An$. The dashed line has unit slope illustrating the Boltzmann scaling, Eq.~\ref{eq:boltz}. The inset shows the logarithmic scaling of $J_a$ with $a$. The dashed line is a fit to the data for Eq.~\ref{eq:logscale} with $g=0.625$.}
\label{fig:meanexcite}
\end{center} 
\end{figure}

The mean mobility (or excitation concentration) is the net number of enduring transitions per molecule per unit time, i.e., 
\begin{equation}
\label{eq:ca}
c_{a} = \langle C_{a} \rangle /(N t_\mathrm{obs}/\Delta t).
\end{equation}
Its dependence upon temperature and displacement length is illustrated in Fig.~\ref{fig:meanexcite}.  According to facilitation theory, $c_a$ should have a Boltzmann temperature dependence, with an energy scale that grows logarithmically with displacement length.  That is,
\begin{equation}\label{eq:boltz}
c_{a} \propto \exp \left [ - J_{a} \left ( 1/T-1/\To \right ) \right ] \,\quad \mathrm{for} \, \quad T<\To \, ,
\end{equation}
and
\begin{equation}\label{eq:logscale}
J_{a'}  = J_{a} - g \, J_{\sigma}\, \ln \left (a'/a\right ) \, ,
\end{equation}
where $\sigma$ is a reference molecular length and $g$ is a system-dependent constant.\footnote{Keys et. al\cite{keys2011excitations} use the symbol $\gamma$ for what we call $g$. We use $\gamma$ to refer to the surface tension.}  The data graphed in Fig. 3 shows that for the model considered, the mW model of water, the theoretical expectations are obeyed.  We have adopted the reference length $\sigma = 2.5 \An$, which is close to the diameter of the molecule in the mW model, and find $g=0.625$, $\To = 244$\,K $\approx \Tx = 250$\,K, and $J_\sigma/\To=23$.

According to facilitation theory\cite{keys2011excitations}, Eqs. \ref{eq:boltz} and \ref{eq:logscale} imply
\begin{equation}\label{eq:east}
\ln( \taul /\tau_\mathrm{o}) = (J_\sigma^2 \, g/d_\mathrm{f}) \left (1/T - 1/\To \right )^2\,\quad  \mathrm{for} \, \quad T<\To \, ,
\end{equation}
where $d_\mathrm{f}$ is the fractal dimension of dynamic heterogeneity, which for $d=3$ is about 2.6.  Equation \ref{eq:east} therefore yields
\begin{equation}
\label{eq:J}
J = J_\sigma\, \sqrt{g/2.3 \, d_\mathrm{f}\,}\,
\end{equation}
where the factor of 2.3 in the square-root accounts for the conversion between base e and base 10 logarithms.\footnote{Keys et. al\cite{keys2011excitations} employ natural logarithms in their use of the parabolic law, and thus the factor of 2.3 does not appear in their equations.}  Applying Eq. \ref{eq:J} with the computed parameters yields $J/\To = 7.4$, in good agreement with the universal empirical value reported in the previous section, that empirical value obtained from fitting data for various water models.  Thus, we have succeeded at deriving this value from a molecular calculation.

\section{Theory for crystallization time}
To estimate the timescale for crystallization, $\tau_\mathrm{xtl}$, we start with the usual form, 
\begin{equation}\label{eq:txtl1}
\tau_\mathrm{xtl} = \nu^{-1}(T) \, \mathrm{e}^{\Delta F(T)\,/T} \,
\end{equation}
where $\Delta F(T)$ is the free energy cost for growing a nascent crystal and $\nu^{-1}(T)$ is the timescale for adding material to the burgeoning phase. Typical forms for $\Delta F(T)$ can be motivated by classical nucleation theory, which has been shown previously to yield accurate results for nucleation rate of models of water at moderate supercooling.\cite{li2011homogeneous} In general, this free energy can be written as,
\begin{equation}\label{eq:txtl2}
\Delta F(T) / T = \Phi(\gamma/\Delta h) (T/ \Tm-1)^{-2}
\end{equation}
where $\gamma$ is surface tension for liquid-crystal coexistence, and $\Phi(\gamma/\Delta h)$ is function of the ratio of that quantity to the enthalpy difference between those phases.  The ratio is approximately temperature independent.\cite{debenedetti1996metastable, limmer2012phase} The temperature-dependent factor, $(T/\Tm - 1)^{-2}$, comes from expanding the chemical potential difference to lowest non-trivial order in $T - \Tm$.  

The timescale for adding material to a growing cluster, $\nu^{-1}(T)$, is expected to be relatively athermal at high temperatures, but to increase with supercooling. We expect $\nu^{-1}(T) \propto D(T)$, where $D$ is the molecular self-diffusion constant. Supercooled liquids generically obey a fractional Stokes-Einstein relationship,\cite{ediger2000spatially} 
\begin{equation}\label{eq:ser}
D \propto \taul^{-z} \, 
\end{equation}
For $T>\To$, $z=1$.\cite{hansen2006theory}  On the other hand, for $T<\To$, $z\approx 3/4$.\cite{ediger2000spatially}  This value for the exponent is predicted by the East model.\cite{jung2004excitation} Adopting a fractional Stokes-Einstein relation with with Eq.~\ref{Eq:parabola} implies a super-Arrhenius form for $\nu^{-1}(T)$,
\begin{equation}\label{eq:txtl3}
\nu^{-1}(T) \propto \exp \left [ 2.3\,z \, J^2 \left (1/T-1/\To \right )^2 \right ] \, .
\end{equation}
By combining Eqs.~\ref{eq:txtl1}--\ref{eq:txtl3}, and $\To \approx \Tx$, we obtain
\begin{equation}\label{eq:scaleform}
\ln(\tau_\mathrm{xtl}/\tau_\mathrm{o}^\mathrm{x}) = \Lambda \left ( 1/\tilde{T} - 1 \right )^2 + \Gamma \, \left (\tilde{T}-\tilde{T}_\mathrm{m} \right)^{-2} \, ,
\end{equation}
where $\Lambda = 2.3\,z\,(J/\To)^2$, $\Gamma = \Phi(\gamma / \Delta h) \, \tilde{T}_\mathrm{m}^2$ and $\tau_\mathrm{o}^\mathrm{x}$ is the proportionality constant in Eq.~\ref{eq:txtl3}.  

\begin{figure}[t]
\begin{center}
\includegraphics[width=8.5cm]{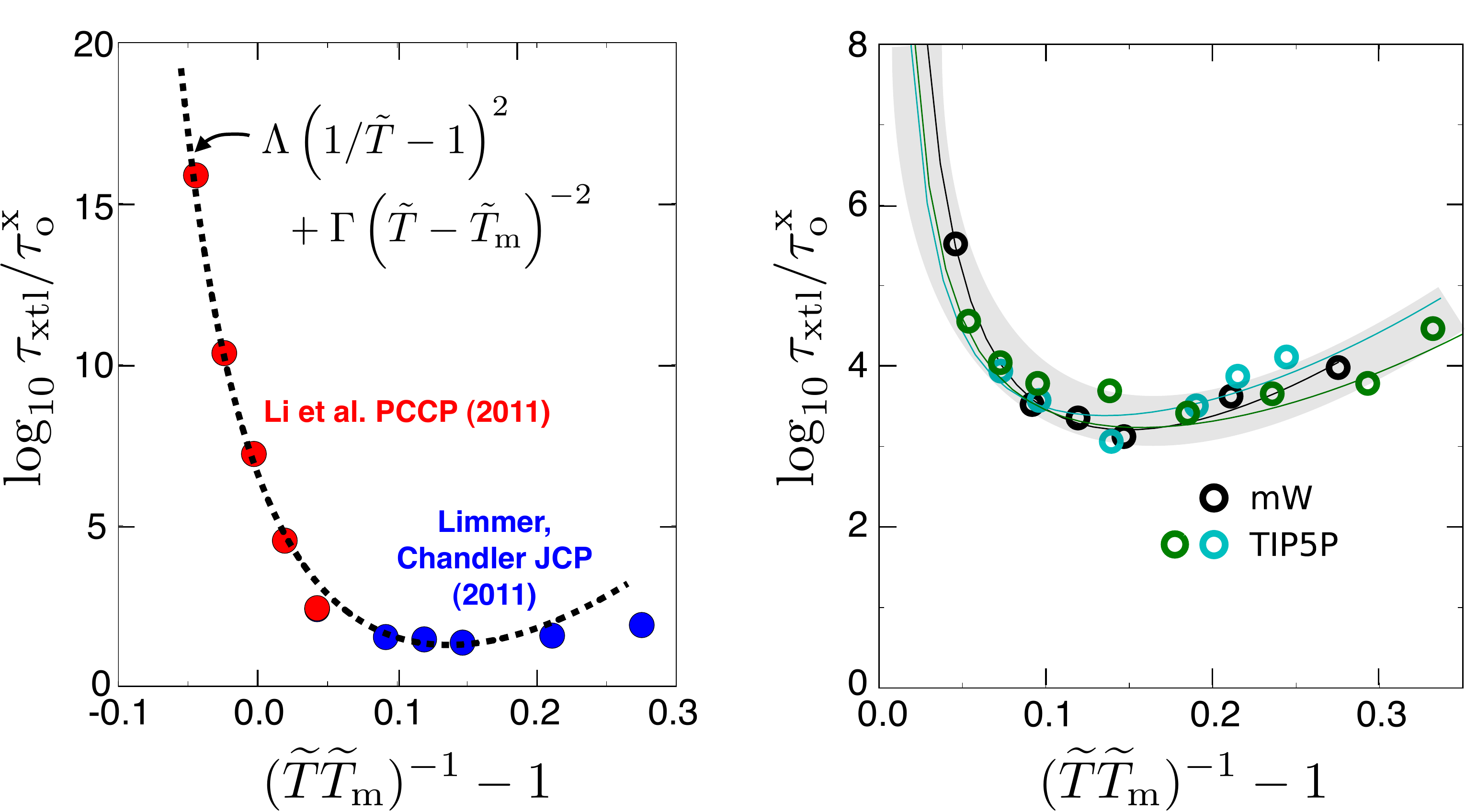}
\caption{Collapse of the crystallization times for models of supercooled water. (\textbf{left}) Crystallization times calculated for the mW model spanning the nucleation and growth regimes. Red markers are taken from Li et. al,\cite{li2011homogeneous} blue markers are taken from Limmer and Chandler.\cite{Limmer:2011p134503} The dashed black line is given by Eq.~\ref{eq:scaleform} with $\Lambda = 94$, $\Gamma = 0.53$ and $\tau_\mathrm{o}^\mathrm{x} = 0.1$. (\textbf{right}) Crystallization times for different models collapse. Black markers are our data for the mW model. Cyan and green markers are for two distinct sets of data for the TIP5P model taken from Yamada et. al.\cite{yamada2002interplay} The three lines are Eq.~\ref{eq:scaleform} with $\Lambda = 94$ and with $\Gamma$ and $\tau_\mathrm{o}^\mathrm{x}$ adjusted for best fits to each of the three different data sets. To within 10\% for each of the three cases, $\Gamma=0.5$ and $\tau_\mathrm{x}=8$ ps.
\label{fig:collapsetxtl}}
\end{center} 
\end{figure}

Equation \ref{eq:scaleform} can be used to fit crystallization rates in terms of the constants $\Lambda$ and $\Gamma$. From the universal East-model value for $z$, and the universal value for $J/\To$ in water-like systems, we have $\Lambda \approx 94$.  Further, by approximating critical nuclei as spherical and mono-disperse,
\begin{equation}\label{eq:cnt}
\Gamma \approx \frac{4 \pi}{3}\left ( \frac{2\gamma}{\Delta h}\right )^2 \frac{\gamma}{ \Tm}  \, .
\end{equation}
For the mW model we have previously determined\cite{limmer2012phase,li2011homogeneous} all the quantities on the right-hand side of Eq.~\ref{eq:cnt}, yielding for that model $\Gamma \approx 0.57$. Equation~\ref{eq:scaleform} is plotted along side the numerical data in Fig.~\ref{fig:collapsetxtl} with this parameterization. The agreement is good over a range of 10 orders of magnitudes, spanning nanoseconds to seconds. The worst agreement is at the lowest temperature, where the rate is the most sensitive to the preparation of the initial state, as the liquid is no longer metastable at this condition.  The next section of this paper expands upon this point.

\section{Time-temperature-transformation diagrams}

At conditions of liquid metastability, where a free energy barrier separates liquid and crystal basins, nucleation is the rate-determining step to form the equilibrium phase. Two data sets taken from the literature have used different rare-event sampling techniques to compute these times for a range of temperatures for the mW model. Limmer and Chandler\cite{Limmer:2011p134503} computed the nucleation rate constant following a standard Bennett-Chandler procedure.\cite{frenkel2001understanding} Li et. al\cite{li2011homogeneous} calculated the nucleation rate using forward-flux sampling\cite{allen2009forward} with an order parameter based on a crystalline cluster sizes. To the extent that the kinetics is nucleation limited, both calculations are expected to have the same temperature dependence. However, because each used different order parameters and basin definitions, the prefactors can be different. In order to compare both data sets, we determine the ratio of prefactors by equating the rate at $T=220$\,K, which was calculated in both studies. These data sets are shown in left panel of Fig. \ref{fig:collapsetxtl}. 

\begin{figure}[t]
\begin{center}
\includegraphics[width=8.5cm]{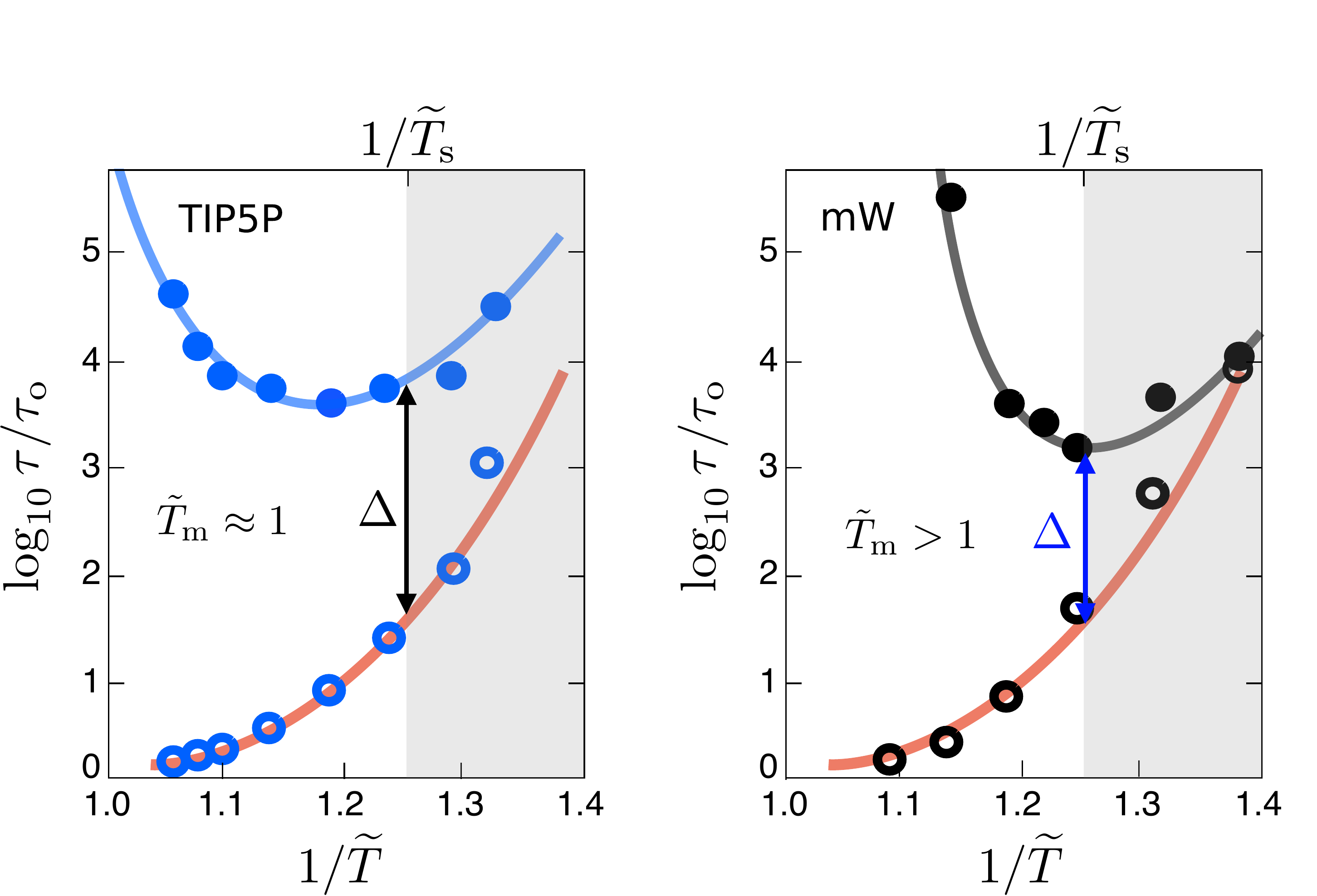}
\caption{Time scales of supercooled TIP5P water (left) and mW water (right) at 1 bar. Open circles are computer simulation results for structural relaxation times, $\tau_\mathrm{liq}$. Filled circles are computer simulation results for crystallization times, $\tau_\mathrm{xtl}$. One standard-deviation error estimates for the mW simulations (our results) are the size of the symbols.  Error estimates for the TIP5P simulations\cite{yamada2002interplay} are unknown.  The red line is the parabolic law, Eq.~\ref{Eq:parabola}, and the blue and black lines are fits to Eq.~\ref{eq:scaleform} with $\Lambda = 94$, as predicted by theory. The grey region is where the liquid is unstable. At the boundary of liquid stability, the separation between times $\tau_\mathrm{liq}$ and  $\tau_\mathrm{xtl}$ defines the gap parameter, $\Delta$.
}\label{Fi:tau_temp}
\end{center} 
\end{figure}

For lower temperatures, we take data sets for crystallization times computed from first-passage calculations.\cite{van1992stochastic} In the calculations of Limmer and Chandler\cite{Limmer:2011p134503}, the first-passage time is taken from simulations with the mW model.  Similar calculations by Yamada et. al\cite{yamada2002interplay} based on first passage times have been calculated for the TIP5P model. In the latter case, the potential energy and structure factor were used as an order parameters for distinguishing crystallization. While Tm for the polymorph that TIP5P freezes into is not known, these data are taken sufficiently far away from any singular response that the fit to eqn 16 is insensitive to its precise value. Both of these calculations are shown in Fig.~\ref{Fi:tau_temp}. 

The crystallization times shown in Figs.~\ref{fig:collapsetxtl} and \ref{Fi:tau_temp} illustrate the non-monotonic temperature dependence predicted from Eq.~\ref{eq:scaleform}. In the higher-temperature regime, nucleation rates increase because the barrier to nucleation decreases in size.  In the lower-temperature regime, the process of crystallization is slowed by the onset of glassy dynamics. At conditions where the amorphous phase is unstable, $\tau_\mathrm{xtl}$ becomes limited by mass diffusion, which from Eq.~\ref{eq:ser} is proportional to $\tau_\mathrm{liq}$. In this region of the phase diagram, the liquid state is no longer physically realizable.  

In plotting $\tau_\mathrm{xtl}$ in Fig. \ref{fig:collapsetxtl}, we use a different reduced temperature scale than previous plots of $\taul$. The different scale is chosen to emphasize the crossover region, where nucleation and growth compete. This particular scale also allows for crystallization times to be collapsed for different models because, to first order in $\Lambda/\Gamma$, this scale locates the minimum in $\tau_\mathrm{xtl}$. The location is the solution to a quadratic polynomial found by equating the nucleation and growth terms.\footnote{The solution for the minima is $\tilde{T} =(\tilde{T}_\mathrm{m}+\omega)/2 +[(\omega+ \tilde{T}_\mathrm{m})^2 - 4 \tilde{T}_\mathrm{m}]^{1/2}/2$, where  $\omega=1-(\Lambda/\Gamma)^{1/2}$.} This scaling holds only for $T$ far below $\Tm$. Away from the singular response at $T=\Tm$, this form is conserved from model to model as it reflects the the crossover to universal structural relaxation times away from the nucleation dominated regime.  

In Fig.~\ref{Fi:tau_temp} we show both $\taul$ and $\tau_\mathrm{xtl}$ to illustrate how the separation of timescales evolves as a function of temperature for different models of water. By considering two cases, where $\To \approx \Tm$ and where $\To < \Tm$, we see large variation in the time-scale gap between liquid relaxation fastest crystallization.  To quantify this variation between models, we define
\begin{equation}\label{eq:Delta}
\Delta=\log_{10} \frac{\tau_\mathrm{xtl}}{\taul} \Big |_{T=\Ts} \, ,
\end{equation}
and subtract Eq.~\ref{Eq:parabola} from \ref{eq:scaleform} to predict how this gap parameter changes with $\tilde{T}_\mathrm{m}$.  For the mW model, the parameters in this equation predict $\Delta \approx 1.6$, in agreement with simulation. For experimental water, $\tilde{T}_\mathrm{m} = 0.99$, and $\Gamma$ can be computed using Eq.~\ref{eq:cnt} and known values for $\gamma$ and $\Delta h$, yielding $\Gamma \approx 0.52$.\footnote{\citet{eisenberg2005structure} gives $\Delta h \approx 3.0 \times 10^5 \,\mathrm{kJ/m^3}$. Granasy et al.\cite{granasy2002interfacial} give a range of values for $\gamma$ from which we take $\gamma \approx 25 \,\mathrm{mJ}/\mathrm{m}^2$.}  As such, Eq.~\ref{eq:Delta} gives $\Delta = 3.4$, consistent with cooling rates required to bypass crystal nucleation.\cite{koop:2000p611}  The location of the minimum crystallization time for experiment can be similarly predicted, and this yields $\tilde{T}=0.77$ or $T\approx 215$~K, which is close to, though lower than, previous estimates.\cite{moore2011structural} One may also use this analysis to predict the time scales on which models of water will exhibit complex coarsening dynamics resulting in artificial polyamorphism.\cite{limmer2013putative}  

\begin{center}
\begingroup
\small  
\LTcapwidth=250pt
\begin{longtable}{lccccccccc}
\caption{\small Summary of properties for water and water models, with standard acronyms identifying different models.\cite{broughton1987phase,eisenberg2005structure,abascal2005general,vega2005relation,Molinero:2009p4008,Stillinger:1974p1545,Stillinger:1985p5262,limmer2012phase}  See text for meanings of symbols and the methods by which the properties are determined. Absolute temperatures are in K, pressures are in kbar and times are in ps.}\\ \\
\hline\hline \\[-1ex]
    \multicolumn{1}{l}{Model\ } &
  \multicolumn{1}{c}{$\Tx$} &
   \multicolumn{1}{c}{$p_\mathrm{o}$} &
  \multicolumn{1}{c}{$\tilde{T}_\mathrm{m} $} &
   \multicolumn{1}{c}{$\tilde{T}_\mathrm{o}$  } &
   \multicolumn{1}{c}{$J/\To$  } &
   \multicolumn{1}{c}{$\tau_\mathrm{o}$} &
   \multicolumn{1}{c}{$\tau_\mathrm{o}^\mathrm{x}$} &
    \multicolumn{1}{c}{$\Gamma$} &
   \multicolumn{1}{c}{$\Delta$} \\[0.5ex] \hline
   \\[-1.9ex]
Experiment &	277 	&	3.7	&	0.99	&	0.98 & 7.4 & 1.0 & 0.3 & 0.52 & 3.4 \\
mW		&	250 	&	10.0	&	1.09	&	0.98	&  7.0 & 0.6 & 0.1 & 0.57 & 1.6 \\
SW		&	1350 	&	16.6	&	1.20	&	-	&	- & - & -&-& - \\
SPC/E	&	241 	&	2.7	&	0.89	&	1.03	&	7.7 & 0.4 & -&-& - \\
ST2		&	320 	&	3.4	&	0.94	&	0.95 &	7.6 & 3.0 & -&-&  2.4 \\
TIP4P	&	253 	&	3.7	&	0.92	&	-	&	- &  - & -& -& - \\
TIP4P/2005	&	277 	&	3.4	&	0.90	&	0.99	& 7.5 & 9.0 & - & - & - \\
TIP5P		&	285 	&	19.4 &	0.96	&	0.98	& 7.6  & 0.2 & 8.0 & 0.50 & 2.1  \\
\hline
\label{table:fragile}
\end{longtable}
\endgroup
\end{center}

Table 1 summarizes materials properties noted in this and preceding sections.  Blanks (-) in the table refer to properties that have not yet been determined. Viewing the variability between models for the values for $\Tx$ and $p_\mathrm{o}$ elucidates how apparent different behaviors of different models can simply reflect different corresponding states.

\section{Mesostructured supercooled water}
The lengthscale over which the arguments presented in the previous sections are applicable to supercooled water reflects the lengthscale over which orientational order is correlated. We have previously studied these correlations using the phenomenological hamiltonian of the form,
\begin{equation}\label{eq:ham}
\mathcal{H}[q(\mathbf{r})] = \int_\mathbf{r} \left( f[q(\mathbf{r})] + \frac{m}{2} | \nabla q(\mathbf{r}) |^2 \right) \, ,
\end{equation}
where $q(\mathbf{r})$ is an order parameter that measures the amount of local orientational order at a point $\mathbf{r}$, $f(q)$ is a free energy density,
\begin{equation}
f(q) = a(T-\Ts) q^2/2 - w q^3 + uq^4\,,
\end{equation}
and $a$, $w$, $u$ and $m$ are positive constants determined by $\Delta h$, $\Tm$ and $\gamma$.\cite{limmer2012phase} This hamiltonian is isomorphic with that of van der Waals for liquid-vapor coexistence.\cite{rowlinson2002molecular} Consequently, mean profiles for $q(\mathbf{r})$ subject to external boundary conditions yield smooth order-parameter profiles like those at a liquid-vapor interface. Instantaneously, this field can be represented in a discrete basis and sampled with an interacting lattice gas.\cite{goldenfeld1992lectures} Such a coarse-grained representation is amenable to large-scale computations, beyond what are tractable with atomistic models.

\begin{figure*}[t]
\begin{center}
\includegraphics[width=13.0cm]{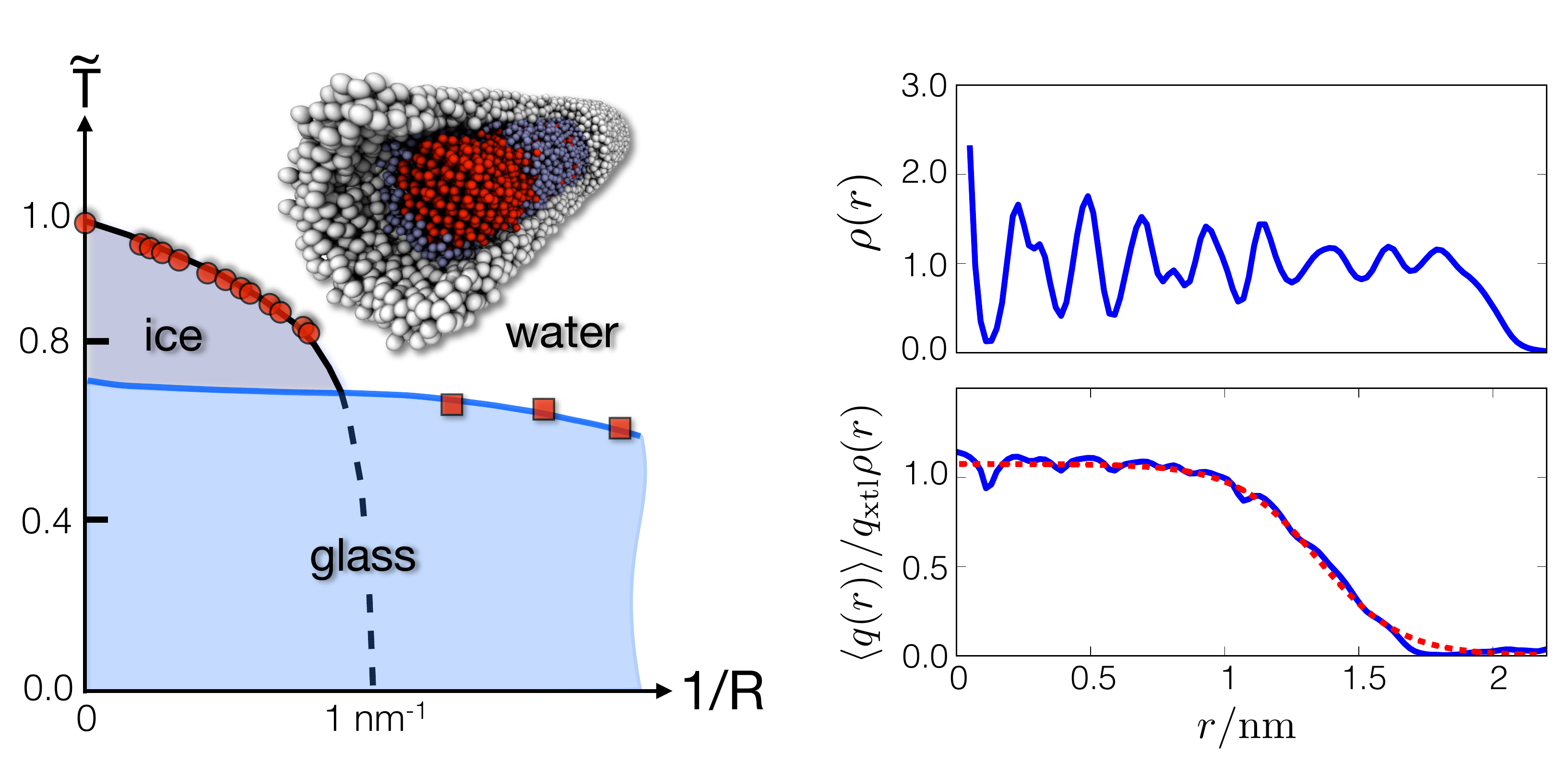}
\caption{Liquid-solid transitions of water confined to hydrophilic nanopores. (\textbf{left}) The phase diagram for water confined to hydrophilic nanopores divides into regions of liquid, glass and crystal-like states. The black line locates $\Tm(R)$ while the blue locates $\Tg(R)$. Red markers are experimental data for the melting line (circles)\cite{findenegg2008freezing} and glass transition line (squares).\cite{oguni2011calorimetric} The top part of this panel illustrates a typical configuration of ice-like water, shown in red, in contact with the hydrophilic nanopore, shown in grey, mediated with a premelting layer, shown in blue.  The configuration is taken from molecular dynamics calculations of the mW model.\cite{limmer2012phase} (\textbf{right}) The mean molecular density (top), and mean local orientational order parameter (bottom), for mW water confined to a $R=20\,\An$ pore for $T<\Tm$, where $q_\mathrm{xtl}$ is the mean value of the order parameter in the center of the pore. The red dashed line in the bottom right panel is the theoretical prediction from the square-gradient theory in Eq.~\ref{eq:ham}.}\label{fig:confine}
\end{center} 
\end{figure*}

One case of water interacting with mesoscopic inhomogeneities that we have studied previously is water confined to hydrophilic nanopores.\cite{limmer2012phase} For nanopores with radii greater than, $R>1$~nm, the properties of the water enclosed in the pore are sufficiently bulk-like that these scaling relations hold up to a perturbation due to the surface. Indeed using the expression in Eq.~\ref{eq:Tg} we have shown that the locations of glass transitions in $\tilde{p}_\mathrm{o}$--$R$ plane can be predicted.\cite{limmer2012phase}  These results are summarized in Fig.~\ref{fig:confine} which shows a $\tilde{p}_\mathrm{o}=0$ cut through the $\tilde{p}_\mathrm{o}$--$R$ plane. The location of the the glass transition, $\Tg$ for finite pores has been measured.\cite{oguni2011calorimetric} These points are included in Fig.~\ref{fig:confine} and fall on our predicted glass transition line.

We have also previously computed the melting temperature in confinement from the partition function prescribed by Eq.~\ref{eq:ham}.\cite{limmer2012phase} The resulting melting temperature as a function of pore size and pressure is given by
\begin{equation}
\label{eq:melt}
T_\mathrm{m}(p,R) =T_\mathrm{m}(p) \left [ 1-\ell_\mathrm{m}/R - \ell_\mathrm{s}^2/8\pi (R-\ell_\mathrm{s}) R\right ]\, 
\end{equation}
where $\ell_\mathrm{m}=2\gamma/\Delta h$ reflects the typical spatial modulations in local order and $\ell_\mathrm{s}=\ell_\mathrm{m}/(1-\Ts/\Tm)$ is the renormalized length that reflects fluctuations that destabilize order. For experimental water, $\ell_\mathrm{m} \approx 2.1$\,\AA, and $\ell_\mathrm{s} \approx 9.1$\,\AA.  This reduction in the melting temperature, Eq.\,\ref{eq:melt}, is a consequence of the silica pore wall stabilizing an adjacent disordered surface of water. The disordered surface shifts the conditions of coexistence. The melting line calculated from this equation is plotted in Fig.~\ref{fig:confine} and compared with the locations of previously determined freezing temperatures for water in silica nanopores.\cite{findenegg2008freezing} As with the glass transition line, there is good agreement with experimental data. In our prior work,\cite{limmer2012phase} we have also used this understanding of the phase diagram to explain the existence of a dynamic crossover and recent observations of hysteresis in density measurements for water confined to MCM-41 silica nanopores.\cite{bertrand2013deeply}

\begin{figure*}[t]
\begin{center}
\includegraphics[width=13.0cm]{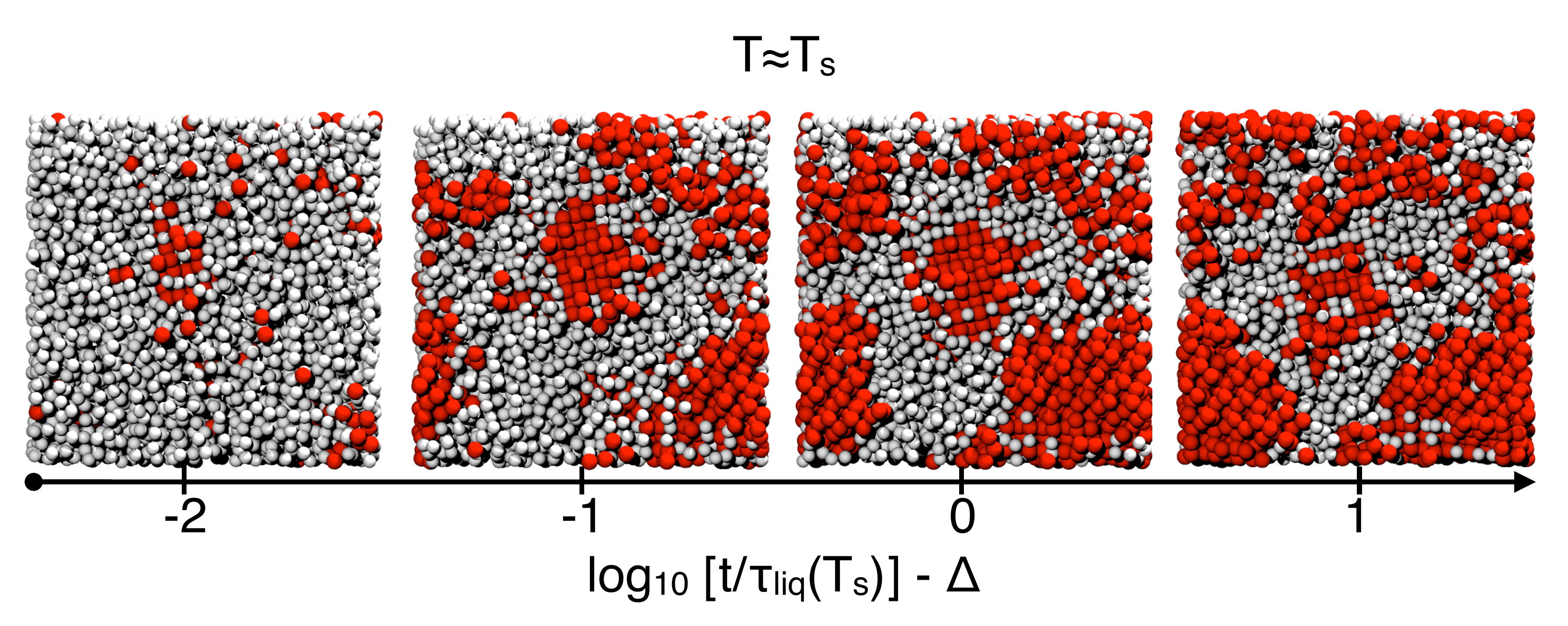}
\caption{Example of a coarsening trajectory from molecular dynamics simulations of the mW model liquid quenched to a temperature $T\approx\Ts$.  The pressure is fixed at $p=0$ throughout the trajectory.  At these conditions the initial liquid configuration is unstable, and the system evolves slowly to the crystal.  The time over which this coarsening occurs is much longer than that of liquid relaxation time, as emphasized by the logarithmic time scale.  Assuming that $\tau_\mathrm{o}^\mathrm{x} \approx \tau_\mathrm{o}$, which is expected to hold within one order of magnitude for the models we have examined, the universal form of the logarithmic scale can be used to predict coarsening time scales for other models and experiment.  Red spheres locate the positions of the molecular centers that are locally crystal-like and grey spheres locate the positions of the molecular centers that are locally liquid-like.  The pictures are from cuts through the simulation box at the times indicated by the tick marks on the time line.  The simulation employs periodic boundary conditions. }\label{fig:coarse}
\end{center} 
\end{figure*}

We mention that explanation here because it relates to another instance of water evolving into mesoscopic structures, specifically the recent experimental observations by Murata and Tanaka.\cite{murata2012liquid}  Complex structure emerges from a mixture of water and gylcerol as it is quenched to low temperatures. The patterns observed depend on the depth of the quench and the relative concentrations of the two components. These patterns are reminiscent of the early stages of coarsening that we have found from theory for pure water near $\Ts$. A specific example of such structural evolution is illustrated in Fig.~\ref{fig:coarse}, where the bulk free energy barrier to crystallization disappears. Nucleation occurs throughout the system and growth becomes the limiting timescale. This behavior is reflected in the gap in timescales between density and long ranged order evolution, as quantified by $\Delta$. Combining the quantitative understanding of timescales developed in this work with the understanding of how ice surfaces are modulated according to the phenomenological hamiltonian in Eq.~\ref{eq:ham} may admit a simple explanation for the observations of Murata and Tanaka.\cite{murata2012liquid} 

\section*{Acknowledgements}
Work on this project was supported by the Helios Solar Energy Research Center, which is supported by the Director, Office of Science, Office of Basic Energy Sciences of the U.S. Department of Energy under Contract No. DE-AC02-05CH11231.

\footnotesize{}
%

\end{document}